\documentclass[a4paper,11pt]{article}
\pdfoutput=1 

\usepackage{jheppub} 

\usepackage[T1]{fontenc} 
\usepackage{comment}
\usepackage{soul}
\usepackage[textsize=footnotesize,textwidth=2.cm]{todonotes}
\usepackage{hyperref}
\usepackage{float}

\def\p{\partial}
\newcommand{\ba}{\begin{array}}
\newcommand{\ea}{\end{array}}
\newcommand{\bi}{\begin{itemize}}
\newcommand{\ei}{\end{itemize}}
\newcommand{\bea}{\begin{eqnarray}}
\newcommand{\eea}{\end{eqnarray}}
\newcommand{\be}{\begin{equation}}
\newcommand{\ee}{\end{equation}}
\newcommand{\nn}{\nonumber}

\title{\boldmath Warped CFT duals of the Pleba\'nski-Demia\'nski family of solutions}

\author[\dagger]{Xuhao Jiang}
\author[*]{and Jianfei Xu}

\affiliation[\dagger]{Technical University of Munich, TUM School of Natural Sciences, Physics Department, 85748 Garching, Germany}
\affiliation[*]{Shing-Tung Yau Center and School of Mathematics, Southeast University, Nanjing, 210000, China}

\emailAdd{xuhao.jiang@tum.de}
\emailAdd{jfxu@seu.edu.cn}

\date{\today}

\abstract{In this paper, we analyze the symmetry properties of  the complete family of type D spacetimes generalized form the Pleba\'nski-Demia\'nski solution in four dimensions holographically in terms of a warped CFT. The generalized Pleba\'nski-Demia\'nski solutions are black hole-like spacetimes characterized by seven physical parameters. Most of the black holes in four dimensions are included within this family. Generically consider a solution with horizon in this family, we figure out the possible warped conformal symmetry attached to the horizon. The horizon can be either extremal or non-extremal. In the extremal case, the near horizon region can be mapped to an infinite spacetime with geometry given by a warped and twist product of AdS$_2$ and S$^2$. The new boundary conditions for AdS$_2$ as well as their higher dimensional uplifts are applied here to manifest the asymptotic symmetry as the warped conformal symmetry. In the non-extremal case, the global warped conformal symmetry is singled out by analyzing the scalar wave equation with constant frequency. The local warped conformal symmetries are represented by the charge algebra associated to the vector fields which preserve the scalar wave equation as well as its frequency. In defining the variation of the covariant charges, a proper counterterm is introduced for consistency conditions which is supposed to be suitable for all the solutions within the family. As a consistency check, the horizon entropy is reproduced by the entropy formula of the warped CFT by using its modular covariance and the central terms derived in the bulk spacetimes.}

\begin{document}
\maketitle
\flushbottom
\section{Introduction}
Symmetries are crucial mathematical structures that underpin numerous physical theories. In the realm of black hole physics, the asymptotic symmetry analysis of spacetime geometries under specific boundary conditions helps to reveal potential dual field theories of black holes. The seminal work in that direction comes with a concrete set up of the duality based on the asymptotic symmetry analysis in three dimensions, which is known as the AdS$_3/$CFT$_2$~\cite{Brown:1986nw}. Utilizing this duality, black holes with AdS factors acquire microscopic descriptions under which their thermal entropy can be reproduced by the asymptotic growth of states in a 2D CFT~\cite{Strominger:1997eq, Maldacena:1998bw}.

Explicit AdS factors are however not necessary conditions for black holes to acquire CFT descriptions. One of the counter example is the Kerr black hole. Kerr black holes are four dimensional asymptotically flat spacetimes with two parameters labeling their mass and angular momentum. For an extremal Kerr black hole with outer and inner horizons coinciding, its near horizon region can be mapped to an infinite large spacetime with $SL(2, R)\times U(1)$ isometry~\cite{Bardeen:1999px}. Under specific boundary conditions, the $U(1)$ isometry can be enhanced to the local conformal symmetries represented by the Virasoro algebra. The central charge in the Virasoro algebra is found to be proportional to the angular momentum of the extremal Kerr black hole. The Cardy formula with the derived central charge then can be used to reproduce the black hole entropy in the extremal case. This duality is known as the Kerr$/$CFT~\cite{Guica:2008mu}. For an non-extremal Kerr black hole, the near horizon region no longer has conformal symmetry as it is approximated by a Rindler space. Nevertheless, the wave equation of scalar perturbations near the horizon of the non-extremal Kerr background can be shown to be identical to the Casimir equation of $SL(2, R)$ with two copies. This $SL(2, R)\times SL(2, R)$ symmetry is known as the hidden conformal symmetry which should be understood as intrinsic to the Kerr black hole and governs the dynamics of the scalar perturbation on it~\cite{Castro:2010fd}. See~\cite{Chen:2010bh} for the relevant discussion of hidden conformal symmetry for more general Kerr-Newmann-AdS-dS black holes. The hidden conformal symmetry is a global symmetry which will be spontaneously broken to two translational symmetries in regarding the periodicity along the angular direction. Interestingly, these translational symmetries can be enhanced to the local conformal symmetries represented by two copies of the Virasoro algebra once two sets of chosen vector fields are taken into account in the near horizon region. The vector fields are in forms that infinitesimally keep the Casimir operator of $SL(2, R)$ invariant. The corresponding central charges in the two copies of the Virasoro algebra can be carried out by considering the Dirac brackets among the vector fields induced linearized covariant charges.~\cite{Haco:2018ske}. In defining the variation of the covariant charges, a Wald-Zoupas counterterm is introduced to guarantee the central terms state independent, which is also applicable to Kerr-Newmann black holes~\cite{Haco:2019ggi}. With the help of the counterterm, the central charges are found to share the same expressions as in the extremal case in terms of the angular momentum and the Cardy formula now with both left and right-moving sectors recovers the black hole entropy in the non-extremal case~\cite{Haco:2018ske, Haco:2019ggi}.

In this paper, we will generalize the symmetry analysis of the Kerr black hole both on the gravity side and on the field theory side. On the gravity side, we will consider a complete family of type D spacetimes in four dimensions, which are generalized from the Pleba\'nski-Demia\'nski solution~\cite{Plebanski:1976gy} by certain transformations and limiting procedures~\cite{Griffiths:2005qp}. These spacetimes are characterized by seven parameters with clear physical meanings. Many special cases, like black holes with electric and magnetic charges, rotation, acceleration, NUT charge, or cosmological constant, are included within this family. On the field theory side, we will consider the possible dual field theories of the family of spacetimes as warped CFTs, which are two dimensional non-relativistic quantum field theories. Much like its 2D CFT cousin, the warped CFT also contains infinitely many local symmetries~\cite{Hofman:2011zj}, featured by one Virasoro algebra and one $U(1)$ Kac-Moody algebra~\cite{Detournay:2012pc}. It was initiated from the investigation of the holography of a large class of geometries with $SL(2, R)\times U(1)$ isometry. One typical example of those geometries is the warped AdS$_3$ where consistent boundary conditions can be imposed to realize the symmetries in a warped CFT~\cite{Anninos:2008fx, Compere:2008cv, Compere:2009zj, Blagojevic:2009ek, Anninos:2010pm, Anninos:2011vd, Henneaux:2011hv}. The complete family of spacetimes generalized from the Pleba\'nski-Demia\'nski solution can be shown to possess $SL(2, R)\times U(1)$ isometry near their extremal horizons which makes it natural to take the warped CFTs as their dual field theories. As a field theory with extended local symmetries, many properties of the warped CFT have been uncovered especially in the context of holography~\cite{Hofman:2014loa, Castro:2015uaa, Detournay:2015ysa, Castro:2015csg, Song:2016gtd, Song:2017czq, Jensen:2017tnb, Apolo:2018eky, Chaturvedi:2018uov, Wen:2018mev, Apolo:2018oqv, Song:2019txa, Chen:2019xpb, Gao:2019vcc, Chen:2020juc, Apolo:2020bld, Apolo:2020qjm, Chen:2022fte, Aggarwal:2022xfd}.

The generalized Pleba\'nski-Demia\'nski solutions are black hole-like spacetimes each with a horizon polynomial being a quartic function of the radius, which indicates that each of the spacetimes contains at most four different kinds of horizons depending on the relation among parameters. Generically picking a solution, we will focus on one of its horizons and analyze the possible warped conformal symmetries attached to the horizon. The horizon is said to be extremal if it is a double zero of the horizon polynomial and non-extremal when it is a single zero. In the extremal case, the near horizon region can be mapped to an infinite large spacetime with an AdS$_2$ factor where specific boundary conditions can be imposed at the asymptotic boundary. To figure out the warped conformal symmetries, the new boundary conditions for AdS$_2$~\cite{Godet:2020xpk} as well as their higher dimensional uplifts~\cite{Detournay:2023zni} will be invoked here to pick out the allowed differomorphisms induced by the consistent asymptotic Killing vectors. The corresponding covariant charges of the asymptotic Killing vectors then can be shown to satisfy the warped conformal algebra with non-trivial central extensions. In the non-extremal case with non-zero horizon temperature, the conformal symmetries somehow hide in the wave equation of perturbations on the fixed background like in the Kerr case. We will consider a perturbative massless scalar field to exhibit the hidden conformal symmetry. The scalar field is conformally coupled to the gravity with its wave equation being separable after a conformal transformation. The separated radial wave equation then can be casted into the Casimir equation of $SL(2, R)$ with two copies. As discussed in~\cite{Xu:2023jex, Xu:2023tfl}, the global warped conformal symmetry as a subset of the hidden conformal symmetry survives under the constraint that the scalar field having fixed constant frequency. Imposing this constraint in the bulk will restrict our discussion to the warped CFT holography. Similar to the Kerr case, the local warped conformal symmetries can be represented by the vector fields with forms infinitesimally keeping the Casimir operator from the survived $SL(2, R)$ generators as well as the $U(1)$ generator of the global warped symmetry invariant. The Dirac brackets among covariant charges associated to the vector fields are found to be the warped conformal algebra with non-trivial central extensions.

In deriving the standard warped confromal algebra on the non-extremal background, a proper counterterm is introduced in the definition of the charge variation. In the non-extremal Kerr black hole case, the counterterm is chosen as the Wald-Zoupas counterterm which is originally proposed in~\cite{Haco:2018ske, Haco:2019ggi} and has been shown also essential in recovering the warped conformal algebra for Kerr black holes~\cite{Aggarwal:2019iay} and accelerating Kerr black holes~\cite{Xu:2023tfl}. The choice of the counterterm is however not unique. For Kerr black holes with cosmological constant or NUT charge, a more general counterterm is need for getting the right Virasoro central charges~\cite{Perry:2020ndy, Perry:2022udk}. In the present paper, the bulk geometry we are considering is a black hole-like spacetime with all possible physical parameters. We define a new kind of counterterm so that the charge algebra agrees with the standard warped conformal algebra with vanishing mixed central extension term in the non-extremal case. The counterterm here works for the complete family of spacetimes where all the previously studied black hole cases are included. The warped conformal algebras derived in the extremal and non-extremal cases both have non-trivial central extensions and they are consistent with each other in the extremal limit. As a consistency check, we will show that the entropy associated to the horizon through the area law can be recovered from the entropy formula of a finite temperature warped CFT, which is a dual field theory calculation using the modular properties of the warped CFT partition as well as the dictionaries of the Virasoro central charge, Kac-Moody level, and thermal parameters obtained in the non-extremal bulk spacetime.

This paper is organized as follows. In section \ref{sec2}, we analyze the basic properties the family of generalized Pleba\'nski-Demia\'nski solutions and carry out the hidden conformal symmetry from scalar perturbations. Keeping the scalar frequency as a constant, the allowed generators of the hidden conformal symmetry will form a subalgebra generating the global warped conformal symmetry. Section \ref{sec3} is for presenting the local warped conformal symmetries in the phase space near the horizon with different methods for the extremal and non-extremal cases. For the extremal case, the near horizon region can be mapped to an infinite scaling region where the new boundary conditions for AdS$_2$ can be applied here at the asymptotic boundary. The asymptotic symmetry is found to be identical to the local symmetries of a warped CFT. For the non-extremal case, the linearized covariant charges associated to the vector fields that preserve the scalar radial equation and frequency satisfy the warped conformal algebra under Dirac brackets. The variation of the covariant charges are defined with a proper counterterm which is essential in obtaining the right central extension terms. In section \ref{sec4}, we perform a consistency check for the validity of the warped conformal symmetries by reproducing the horizon entropy of the black hole-like spacetime from the entropy formula of a finite temperature warped CFT. Section \ref{sec5} is for summary and discussion.

\section{Scalar perturbations and hidden conformal symmetry}\label{sec2}
The metric of the black hole-like spacetime generalized from the Pleba\'nski-Demia\'nski solution~\cite{Plebanski:1976gy} can be written as the following form in the Boyer-Lindquist coordinate system~\cite{Griffiths:2005qp}
\begin{align}\label{GPDmetric}
ds^2=\frac{1}{\Omega^2}\Bigg\{&-\frac{Q}{\rho^2}\left[dt-\left(a(1-x^2)+2l(1+x)\right)d\phi\right]^2+\rho^2\left[\frac{dr^2}{Q}+\frac{dx^2}{(1-x^2)P}\right]\nn\\
&+\frac{a^2(1-x^2)P}{\rho^2}\left[dt-\frac{r^2+(a+l)^2}{a}d\phi\right]^2\Bigg\}\,,
\end{align}
where
\begin{align}
\Omega&=1-\frac{\alpha}{\lambda}(l-ax)r\,,\nn\\
\rho^2&=r^2+(l-ax)^2\,,\nn\\
P&=1+a_3x+a_4x^2\,,\nn\\
Q&=(\lambda^2k+e^2+g^2)-2Mr+\epsilon r^2-2\frac{\alpha n}{\lambda}r^3-\left(\alpha^2k+\frac{\Lambda}{3}\right)r^4\,,\nn\\
a_3&=2\frac{\alpha Ma}{\lambda}-4\frac{\alpha^2al}{\lambda^2}(\lambda^2k+e^2+g^2)-4al\frac{\Lambda}{3}\,,\label{mpol}\\
a_4&=\frac{\alpha^2a^2}{\lambda^2}(\lambda^2k+e^2+g^2)+a^2\frac{\Lambda}{3}\,,\nn\\
k&=\left[1+2\frac{\alpha Ml}{\lambda}-3\frac{\alpha^2l^2}{\lambda^2}(e^2+g^2)-l^2\Lambda\right]\Big/\left(\frac{\lambda^2}{a^2-l^2}+3\alpha^2l^2\right)\,,\nn\\
\epsilon&=\frac{\lambda^2k}{a^2-l^2}+4\frac{\alpha Ml}{\lambda}-(a^2+3l^2)\left[\frac{\alpha^2}{\lambda^2}(\lambda^2k+e^2+g^2)+\frac{\Lambda}{3}\right]\,,\nn\\
n&=\frac{\lambda^2kl}{a^2-l^2}-\frac{\alpha M(a^2-l^2)}{\lambda}+(a^2-l^2)l\left[\frac{\alpha^2}{\lambda^2}(\lambda^2k+e^2+g^2)+\frac{\Lambda}{3}\right]\,.\nn
\end{align}
$t$ and $r$ are the time and radial coordinates and the longitudinal part of the spacetime is labeled by coordinates $x$ and $\phi$ with ranges $x\in[-1, 1]$ and $\phi\in[0, 2\pi)$. There are seven independent physical parameters which are the black hole mass $M$, electric charge $e$, magnetic charge $g$, rotation parameter $a$, acceleration $\alpha$, NUT parameter $l$, and cosmological constant $\Lambda$. The free parameter $\lambda$ can be set to any convenient value if $a$ and $l$ are not both zero. The electric and magnetic charges induce an aligned non-null electromagnetic field $F=dA$ with the vector potential given by
\be
A=\frac{1}{\rho^2}\left\{-er\left[dt-\left(a(1-x^2)+2l(1+x)\right)d\phi\right]+gx\left[adt-(r^2+(a+l)^2)d\phi\right]\right\}\,.
\ee

The curvature singularity of the metric \eqref{GPDmetric} locates at $\rho^2=0$. This occurs at $r=0$ and $x=l/a$ when $|l|\leq|a|$. When $|l|>|a|$, $\rho^2$ can not be zero and the corresponding spacetime is non-singular. The conformal infinity of the metric \eqref{GPDmetric} lies at $\Omega=0$ where the boundary value of the radial coordinate $r$ is determined given the relation between $l$ and $a$. When there is no curvature singularity, $r$ in principle can extend to negative values. We summarize the different ranges of $r$ for different cases with the following table in which we assume $a>0$, $\alpha>0$, and $\lambda>0$ for example
\begin{table}[H]
  \centering
  \begin{tabular}{|c|c|c|c|}
  \hline
  $l<-a$ & \multicolumn{2}{|c|}{$-a\leq l\leq a$} & $l>a$ \\ \hline
  $-1\leq x\leq1$ & $-1\leq x\leq\frac{l}{a}$ & $\frac{l}{a}<x\leq1$ & $-1\leq x\leq1$ \\ \hline
  $r_{\scriptscriptstyle\Omega=0}=\frac{\lambda}{\alpha(l-ax)}<0$ & $r_{\scriptscriptstyle\Omega=0}=\frac{\lambda}{\alpha(l-ax)}\geq0$ & $r_{\scriptscriptstyle\Omega=0}=\frac{\lambda}{\alpha(l-ax)}<0$ & $r_{\scriptscriptstyle\Omega=0}=\frac{\lambda}{\alpha(l-ax)}>0$ \\ \hline
  $r\in\left(\frac{\lambda}{\alpha(l-ax)}, +\infty\right)$ & $r\in\left(0, \frac{\lambda}{\alpha(l-ax)}\right)$ & $r\in(0, +\infty)$ & $r\in\left(-\infty, \frac{\lambda}{\alpha(l-ax)}\right)$ \\
  \hline
  \end{tabular}
  \caption{Ranges of radial coordinate for different $x$ and parameter relations.}\label{rangesr}
\end{table}

The metric \eqref{GPDmetric} has conical singularities at poles where $x=\pm1$. Near $x=-1$, the constant $t$, $r$, and $x$ lines are small spatial circles with the ratio of circumference to radius given by
\be
\lim_{x\to-1}\frac{2\pi}{1-x^2}\sqrt{\frac{g_{\phi\phi}}{g_{xx}}}=2\pi(1-a_3+a_4)\,.
\ee
This corresponds to a conical singularity with a deficit angle of
\be
2\pi(a_3-a_4)=2\frac{\alpha Ma}{\lambda}-\frac{\alpha^2a(4l+a)}{\lambda^2}(\lambda^2k+e^2+g^2)-a(4l+a)\frac{\Lambda}{3}\,.
\ee
Physically, the acceleration of the spacetime describes a pair of black holes constantly accelerating away from each other, and the deficit angle near $x=-1$ is right induced by two cosmic strings connecting those two black holes to infinity which cause the acceleration. Near $x=1$, the constant $t$, $r$, and $x$ lines include closed time-like curves in the stationary region where $Q>0$ due to the non-vanishing NUT parameter $l$. This leads to an additional "torsion" singularity which can be transformed to the other pole by transformation $t'=t-4l\phi$~\cite{Griffiths:2005se}. The constant $t'$, $r$, and $x$ lines near $x=1$ now become space-like circles, and the ratio of circumference to radius can be figured out
\be
\lim_{x\to1}\frac{2\pi}{1-x^2}\sqrt{\frac{16l^2g_{tt}+8lg_{t\phi}+g_{\phi\phi}}{g_{xx}}}=2\pi(1+a_3+a_4)\,.
\ee
This corresponds to a conical singularity with an excess angle of
\be
2\pi(a_3+a_4)=2\frac{\alpha Ma}{\lambda}-\frac{\alpha^2a(4l-a)}{\lambda^2}(\lambda^2k+e^2+g^2)-a(4l-a)\frac{\Lambda}{3}\,.
\ee
The excess angle near $x=1$ can be understood as induced by a strut connecting those two black holes which also cause the acceleration. Only one of the conical singularities at the poles can be removed by a scaling of the angular coordinate $\phi$, and the remaining conical singularity with deficit or excess angle is responsible for the acceleration.

$Q$ is the horizon polynomial whose zeros determine the radius of horizons. For the black hole-like spacetime \eqref{GPDmetric} obtained form the Pleba\'nski-Demia\'nski solution with seven parameters, the horizon polynomial $Q$ in \eqref{mpol} is a quartic function of $r$. Generically, roots of $Q$ can be labeled as
\be
Q=-\left(\alpha^2k+\frac{\Lambda}{3}\right)(r-r_0)(r-r_1)(r-r_2)(r-r_3)\,.
\ee
When $\alpha^2k+\frac{1}{3}\Lambda<0$, the spacetime does not contain a cosmic horizon, the largest root of $Q$ has positive slop which corresponds an event horizon of the black hole. Otherwise, the largest root of $Q$ has negative slop which sets a cosmic horizon for the spacetime. In the following discussion, we will take one of the horizons into account and denote the horizon radius as $r_+$. It can be any of the event horizon, Cauchy horizon, or cosmic horizon. In the near horizon region, the horizon polynomial $Q$ can be expanded up to quadratic order around $r_+$
\bea
Q&\approx&Q(r_+)+\frac{dQ}{dr}\Big|_{r=r_+}(r-r_+)+\frac{1}{2}\frac{d^2Q}{dr^2}\Big|_{r=r_+}(r-r_+)^2+\mathcal{O}\left((r-r_+)^3\right)\nn\\
&\approx&k_+(r-r_+)(r-r_s)+\mathcal{O}\left((r-r_+)^3\right)\,,\label{appQ}
\eea
where
\bea
k_+&=&\frac{1}{2}\frac{d^2Q}{dr^2}\Big|_{r=r_+}\nn\\
&=&\epsilon-6r_+\frac{\alpha n}{\lambda}-6r_+^2(\alpha^2k+\frac{\Lambda}{3})\,,\\
r_s&=&r_+-\frac{1}{k_+}\frac{dQ}{dr}\Big|_{r=r_+}\nn\\
&=&r_+-\frac{1}{k_+}\left[2r_+\epsilon-2M-6r_+^2\frac{\alpha n}{\lambda}-4r_+^3\left(\alpha^2k+\frac{\Lambda}{3}\right)\right]\,.
\eea
Here $r_s$ is not a horizon radius unless there are only two horizons exist. For instance, in the Kerr-Newman case where $\lambda=a$ and $\alpha=l=\Lambda=0$, $r_s=2M-r_+$ is the other horizon radius. The horizon surface gravity $\kappa_H$ and area $A_H$ determine the temperature and entropy of the horizon
\begin{align}
T_H&=\frac{\kappa_H}{2\pi}=\frac{k_+(r_+-r_s)}{4\pi(r_+^2+(a+l)^2)}\,,\label{TH}\\
S_H&=\frac{A_H}{4}=\frac{\pi\lambda^2(r_+^2+(a+l)^2)}{(\lambda-r_+\alpha(a+l))(\lambda+r_+\alpha(a-l))}\,.\label{SH}
\end{align}
The higher than quadratic order terms around $r_+$ in $Q$ does not contribute to the temperature. When $r_+=r_s$, the temperature vanishes and the horizon becomes an extremal one. Later we will see that the quadratic order in $Q$ is also responsible for the hidden conformal symmetry of scalar perturbations in the near horizon region.

Now let us discuss scalar perturbations on the background of the black hole-like spacetime \eqref{GPDmetric} from which the hidden conformal symmetry will become manifest in the low frequency and near horizon limit. Consider a perturbative massless neutral scalar field $\Phi$ conformally coupled to the gravity, and its equation of motion is the conformally coupled Klein-Gordon equation
\be
(\nabla_{\mu}\nabla^{\mu}-\chi R)\Phi=0\,,
\ee
where $R$ is the curvature scalar and $\chi=1/6$ is the coupling constant in four dimensions. The scalar field is not separable due to the conformal factor $\Omega^{-2}$ in the metric. Therefor, we can perform a Weyl transformation $\tilde{g}_{\mu\nu}=\Omega^2g_{\mu\nu}$ so that the Weyl transformed scalar field $\tilde{\Phi}=\Omega^{-1}\Phi$ now become separable and the corresponding field equation for $\tilde{\Phi}$ takes a similar form
\be\label{wKG}
(\tilde{\nabla}_{\mu}\tilde{\nabla}^{\mu}-\chi\tilde{R})\tilde{\Phi}=0\,,
\ee
where the covariant derivative and scalar curvature with tilde are evaluated on
\begin{align}
d\tilde{s}^2=&-\frac{Q}{\rho^2}\left[dt-\left(a(1-x^2)+2l(1+x)\right)d\phi\right]^2+\rho^2\left[\frac{dr^2}{Q}+\frac{dx^2}{(1-x^2)P}\right]\nn\\
&+\frac{a^2(1-x^2)P}{\rho^2}\left[dt-\frac{r^2+(a+l)^2}{a}d\phi\right]^2\,.
\end{align}
The Weyl transformed scalar field $\tilde{\Phi}$ is separable which can be written in the following form
\be
\tilde{\Phi}=e^{-i\omega t+im\phi}R(r)S(x)\,.
\ee
The $t$ and $\phi$ dependence is expressed using their Fourier modes since $\p_t$ and $\p_{\phi}$ are Killing vectors. $\omega$ is the frequency of the scalar field and $m$ is its angular momentum which takes integer values due to the $2\pi$ periodicity along $\phi$. Given this ansatz, the field equation \eqref{wKG} can be separated into radial and angular equations
\be\label{re}
\frac{d}{dr}\left(Q\frac{dR(r)}{dr}\right)+\left[\frac{\left((r^2+(a+l)^2)\omega-am\right)^2}{Q}+\frac{Q''}{6}\right]R(r)=\mathcal{K}R(r)\,,
\ee
\begin{align}\label{ae}
\frac{d}{dx}\left((1-x^2)P\frac{dS(x)}{dx}\right)&-\Bigg[\frac{\left(\left(a(1-x^2)+2l(1+x)\right)\omega-m\right)^2}{(1-x^2)P}\nn\\
&+\frac{2P+4xP'-(1-x^2)P''}{6}\Bigg]S(x)=-\mathcal{K}S(x)\,,
\end{align}
where $\mathcal{K}$ is the separation constant. Both equations above are Heun-type second order ordinary differential equations. In the near horizon region, using the expansion \eqref{appQ}, the radial equation \eqref{re} can be approximately written as
\begin{align}
&\frac{d}{dr}\left((r-r_+)(r-r_s)\frac{dR(r)}{dr}\right)+\Bigg[\frac{\left((r_+^2+(a+l)^2)\omega-am\right)^2}{(r-r_+)(r_+-r_s)k_+^2}-\frac{\left((r_s^2+(a+l)^2)\omega-am\right)^2}{(r-r_s)(r_+-r_s)k_+^2}\nn\\
&+\left(r^2+(r_++r_s)r+r_+^2+r_s^2+r_+r_s+2(a+l)^2\right)\frac{\omega^2}{k_+^2}-\frac{2am\omega}{k_+^2}+\frac{1}{3k_+}+\mathcal{O}(r-r_+)\Bigg]R(r)\nn\\
&=\frac{\mathcal{K}}{k_+}R(r)\,.
\end{align}
To figure out the hidden conformal symmetry, we further assume the frequency $\omega$ to be relatively small so that the near horizon region is within the near region defined by
\be
r\ll\frac{1}{\omega}\,.
\ee
In these limits, $r\omega$ is well approximated by $r_+\omega$, and the potential term of the above radial equation only contains two regular singular points around $r_+$ and $r_s$. Thus, the radial equation can be further reduced to
\begin{align}\label{rere}
&\frac{d}{dr}\left(\Delta\frac{dR(r)}{dr}\right)+\left[\frac{\left((r_+^2+(a+l)^2)\omega-am\right)^2}{(r-r_+)(r_+-r_s)k_+^2}-\frac{\left((r_s^2+(a+l)^2)\omega-am\right)^2}{(r-r_s)(r_+-r_s)k_+^2}\right]R(r)\nn\\
&=\mathcal{K}'R(r)\,,
\end{align}
where $\Delta=(r-r_+)(r-r_s)$ and $\mathcal{K}'=\mathcal{K}/k_+-1/(3k_+)+\mathcal{O}\left((r-r_+),\omega\right)$. This reduced radial equation \eqref{rere} is a hypergeometric equation with solutions transforming in the representation of $SL(2, R)\times SL(2, R)$ which coincides the global conformal symmetry in two dimensions. The equation \eqref{rere} has two regular singular points at $r=r_+$ and $r=r_s$ and the solutions have branch cuts at these points. Around the branch cuts, the corresponding radial solutions take the following forms
\be
R(r)\sim c^+_{out}(r-r_+)^{i\alpha_+}+c^+_{in}(r-r_+)^{-i\alpha_+},~~~~R(r)\sim c^s_{out}(r-r_s)^{i\alpha_s}+c^s_{in}(r-r_s)^{-i\alpha_s}\,,
\ee
where
\be\label{mono}
\alpha_+=\frac{\left(r_+^2+(a+l)^2\right)\omega-am}{k_+(r_+-r_s)},~~~~\alpha_s=\frac{\left(r_s^2+(a+l)^2\right)\omega-am}{k_+(r_+-r_s)}\,,
\ee
are the monodromies. The coefficients $c^+_{out, in}$ and $c^s_{out, in}$ combine the two linearly independent solutions with outgoing and ingoing modes near the surfaces at $r=r_+$ and $r=r_s$, respectively.

To explicitly show the hidden conformal symmetry, it is convenient to invoke the following conformal coordinates
\begin{align}
\omega^+&=\sqrt{\frac{r-r_+}{r-r_s}}e^{2\pi T_R\phi+2n_Rt}\,,\nn\\
\omega^-&=\sqrt{\frac{r-r_+}{r-r_s}}e^{2\pi T_L\phi+2n_Lt}\,,\label{cc}\\
y&=\sqrt{\frac{r_+-r_s}{r-r_s}}e^{\pi(T_R+T_L)\phi+(n_R+n_L)t}\,,\nn
\end{align}
where
\be\label{TLTR}
T_R=\frac{k_+(r_+-r_s)}{4\pi a},~~~T_L=\frac{k_+(r_+^2+r_s^2+2(a+l)^2)}{4\pi a(r_++r_s)},~~~n_R=0,~~~n_L=-\frac{k_+}{2(r_++r_s)}\,.
\ee
$T_R$ and $T_L$ are the dual field theory thermal parameters as we will explain in section \ref{sec4}. In terms of the conformal coordinates, one can define two sets of local conformal vectors with the following forms~\cite{Castro:2010fd}
\begin{align}
H_+&=i\frac{\p}{\p\omega^+}\,,\nn\\
H_0&=i\left(\omega^+\frac{\p}{\p\omega^+}+\frac{y}{2}\frac{\p}{\p y}\right)\,,\label{Hp0m}\\
H_-&=i\left((\omega^+)^2\frac{\p}{\p\omega^+}+\omega^+y\frac{\p}{\p y}-y^2\frac{\p}{\p\omega^-}\right)\,,\nn
\end{align}
and
\begin{align}
\bar{H}_+&=i\frac{\p}{\p\omega^-}\,,\nn\\
\bar{H}_0&=i\left(\omega^-\frac{\p}{\p\omega^-}+\frac{y}{2}\frac{\p}{\p y}\right)\,,\label{bHp0m}\\
\bar{H}_-&=i\left((\omega^-)^2\frac{\p}{\p\omega^-}+\omega^-y\frac{\p}{\p y}-y^2\frac{\p}{\p\omega^+}\right)\,.\nn
\end{align}
The commutation relations among the above conformal vectors satisfy the $\mathfrak{sl}(2, R)\times\mathfrak{sl}(2, R)$ algebra
\begin{align}
[H_0, H_{\pm}]&=\mp iH_{\pm},~~~~[H_-, H_+]=-2iH_0\,,\\
[\bar{H}_0, \bar{H}_{\pm}]&=\mp i\bar{H}_{\pm},~~~~[\bar{H}_-, \bar{H}_+]=-2i\bar{H}_0\,.
\end{align}
In terms of the Boyer-Lindquist coordinates, the local conformal vectors can be expressed as
\begin{align}
H_+&=ie^{-2\pi T_R\phi-2n_R t}\left(\sqrt{\Delta}\frac{\p}{\p r}-\frac{n_L\Delta'+n_R(r_+-r_s)}{4\pi\sqrt{\Delta}(n_RT_L-n_LT_R)}\frac{\p}{\p\phi}+\frac{T_L\Delta'+T_R(r_+-r_s)}{4\sqrt{\Delta}(n_RT_L-n_LT_R)}\frac{\p}{\p t}\right)\,,\nn\\
H_0&=i\left(-\frac{n_L}{2\pi(n_RT_L-n_LT_R)}\frac{\p}{\p\phi}+\frac{T_L}{2(n_RT_L-n_LT_R)}\frac{\p}{\p t}\right)\,,\label{BLHp0m}\\
H_-&=ie^{2\pi T_R\phi+2n_R t}\left(-\sqrt{\Delta}\frac{\p}{\p r}-\frac{n_L\Delta'+n_R(r_+-r_s)}{4\pi\sqrt{\Delta}(n_RT_L-n_LT_R)}\frac{\p}{\p\phi}+\frac{T_L\Delta'+T_R(r_+-r_s)}{4\sqrt{\Delta}(n_RT_L-n_LT_R)}\frac{\p}{\p t}\right)\,,\nn
\end{align}
and
\begin{align}
\bar{H}_+&=ie^{-2\pi T_L\phi-2n_L t}\left(\sqrt{\Delta}\frac{\p}{\p r}+\frac{n_R\Delta'+n_L(r_+-r_s)}{4\pi\sqrt{\Delta}(n_RT_L-n_LT_R)}\frac{\p}{\p\phi}-\frac{T_R\Delta'+T_L(r_+-r_s)}{4\sqrt{\Delta}(n_RT_L-n_LT_R)}\frac{\p}{\p t}\right)\,,\nn\\
\bar{H}_0&=i\left(\frac{n_R}{2\pi(n_RT_L-n_LT_R)}\frac{\p}{\p\phi}-\frac{T_R}{2(n_RT_L-n_LT_R)}\frac{\p}{\p t}\right)\,,\label{BLbHp0m}\\
\bar{H}_-&=ie^{2\pi T_L\phi+2n_L t}\left(-\sqrt{\Delta}\frac{\p}{\p r}+\frac{n_R\Delta'+n_L(r_+-r_s)}{4\pi\sqrt{\Delta}(n_RT_L-n_LT_R)}\frac{\p}{\p\phi}-\frac{T_R\Delta'+T_L(r_+-r_s)}{4\sqrt{\Delta}(n_RT_L-n_LT_R)}\frac{\p}{\p t}\right)\,.\nn
\end{align}
Using the parameters in \eqref{TLTR}, one can verify that the reduced radial equation \eqref{rere} is identical to the following Casimir equation of $SL(2, R)$
\be\label{Casimireq}
\mathcal{H}^2\Psi=h(h+1)\Psi\,,
\ee
where
\begin{align}
\mathcal{H}^2&=\bar{\mathcal{H}}^2=-H_0^2+\frac{1}{2}(H_+H_-+H_-H_+)\nn\\
&=\frac{1}{4}\left(y^2\frac{\p^2}{\p y^2}-y\frac{\p}{\p y}\right)+y^2\frac{\p^2}{\p\omega^+\p\omega^-}\,,\label{Casimir}
\end{align}
is the conformal Casimir operator and the eigenvalue in \eqref{rere} has been set to $\mathcal{K}'=h(h+1)$. $\Psi=e^{-i\omega t+im\phi}R(r)$ is the $x$ independent part of the scalar field which is now endowed with equal left and right-moving conformal weights $(h, h)$. In this sense, the solutions of scalar perturbation equation on the background $\eqref{GPDmetric}$ locally form the $SL(2, R)\times SL(2, R)$ representations. Similar to the Kerr case~\cite{Castro:2010fd}, this is the hidden conformal symmetry intrinsic to the Pleba\'nski-Demia\'nski background and will govern the dynamics of the scalar field on it.

In this paper, we will focus on the warped conformal symmetries appear in the near horizon region. Therefore, not all the conformal generators in \eqref{Hp0m} and \eqref{bHp0m} are legal under this consideration. In the Boyer-Lindquist coordinates, since $n_R=0$ given in \eqref{TLTR}, we have $\bar{H}_0\propto\p_t$ as expressed in \eqref{BLbHp0m}. So the eigenvalue of the generator $\bar{H}_0$ acting on the scalar field is proportional to its frequency $\omega$. However, the actions from $\bar{H}_{\pm}$ generators will change the eigenvalue as well as the scalar frequency since they have non-trivial commutation relations with $\bar{H}_0$. As discussed in~\cite{Xu:2023jex, Xu:2023tfl}, a consistent holographic setup for warped CFT operators requires the corresponding dual bulk fields having fixed constant frequency instead of a variable conjugate to time. In this sense, only $H_{0, \pm}$ and $\bar{H}_0$ are allowed when taking the warped conformal symmetries in the near horizon region into account. $H_{\pm}$ will generate the $SL(2, R)$ symmetry while $\bar{H}_0$ is responsible for the $U(1)$ symmetry. These survived generators form a subgroup $SL(2, R)\times U(1)$ which is identical to the global warped conformal symmetry of a warped CFT.

The global warped conformal symmetry $SL(2, R)\times U(1)$ is spontaneously broken to $U(1)\times U(1)$ by the $2\pi$ periodicity along $\phi$, i.e., $\phi\sim\phi+2\pi$. Under this periodic identification, the conformal coordinates defined in \eqref{cc} inherit the following identifications
\be\label{ccids}
\omega^+\sim e^{4\pi^2T_R}\omega^+,~~~~\omega^-\sim e^{4\pi^2T_L}\omega^-,~~~~y\sim e^{2\pi^2(T_R+T_L)}y\,.
\ee
Only $H_0$ and $\bar{H}_0$ are invariant under the above transformations. So from the hidden symmetry perspective, only the $U(1)\times U(1)$ part is globally well defined and this part will play the role as the translational symmetry along the two coordinates defined in a warped CFT.

\section{Warped conformal symmetries in phase space}\label{sec3}
Nevertheless, in a warped CFT, there are infinitely many extended local symmetries featured by the Virasoro and $U(1)$ Kac-Moody algebra. To support the warped CFT dual of the black hole-like spacetime \eqref{GPDmetric}, one need to further specify the extended local warped conformal symmetries as well in the bulk geometry. In this section, we will present two different methods for recovering the local symmetries in the bulk spacetime \eqref{GPDmetric}. In the extremal case, that is the horizon radius $r_+$ is a double zero of the horizon polynomial $Q$, the near horizon region can be mapped to an infinite scaling region where specific boundary conditions can be imposed to its asymptotic boundary. Then the charge algebra with central extensions defined in the phase space of the asymptotic Killing vectors induced by the boundary conditions can be implemented to include the local warped conformal symmetries. In the non-extremal case, there is no near horizon infinite scaling region. However, one can define a set of local vector fields whose induced infinitesimal diffeomorphisms can keep the Casimir operator $\mathcal{H}^2$ \eqref{Casimir} as well as the $U(1)$ generator $\bar{H}_0$ \eqref{bHp0m} invariant. Such vector fields which will keep the scalar wave equation and frequency invariant in the near horizon region can be viewed as carrying local symmetries represented from both the dynamics and kinematics of the scalar perturbation. The charge algebra with central extensions in the phase space of these vector fields can be shown to represent the local warped conformal symmetries.

\subsection{Extremal case}
In the extremal case, the temperature \eqref{TH} vanishes and the horizon radius $r_+$ coincides with $r_s$ which makes it as a double zero of the horizon polynomial $Q$. Then in the near horizon region where $r$ is close to $r_+$, the divergent behavior in the metric components can be removed by infinite coordinate scalings, and the resultant scaling independent metric characterizes a decoupled near horizon geometry. Explicitly, the near horizon geometry can be obtained by the following coordinate transformation
\be\label{nhct}
r=r_++\delta r_0\tilde{r},~~~~t=\frac{r_0}{\delta}\tilde{t},~~~~\phi=\tilde{\phi}+\Omega_H\frac{r_0}{\delta}\tilde{t}\,,
\ee
where $\delta\to0$ is the near horizon limit, $r_0$ is introduced to cancel redundant factors, and $\Omega_H$ is the horizon angular velocity. For the black hole-like spacetime \eqref{GPDmetric},
\be\label{OH}
\Omega_H=\frac{a}{r_+^2+(a+l)^2}\,,
\ee
and the near horizon geometry in the extremal case where $r_+=r_s$ can be expressed as
\be\label{nhmetric}
ds^2=\Gamma(x)\left(-\tilde{r}^2d\tilde{t}^2+\frac{1}{\tilde{r}^2}d\tilde{r}^2+\sigma^2(x)dx^2+\gamma^2(x)(d\tilde{\phi}+b\tilde{r}d\tilde{t})^2\right)\,,
\ee
where
\begin{align}
r_0&=\sqrt{\frac{r_+^2+(a+l)^2}{k_+}},~~~~\Gamma(x)=\frac{\lambda^2(r_+^2+(l-ax)^2)}{k_+(\lambda-r_+\alpha(l-ax))^2},~~~~\sigma^2(x)=\frac{k_+}{(1-x^2)P}\,,\\
b&=\frac{2r_+a}{k_+(r_+^2+(a+l)^2)},~~~~\gamma^2(x)=\frac{k_+(r_+^2+(a+l)^2)^2(1-x^2)P}{(r_+^2+(l-ax)^2)^2}\,.
\end{align}
The $\delta$ independent near horizon metric \eqref{nhmetric} has $SL(2, R)\times U(1)$ isometry and can be viewed as a warped and twisted product of AdS$_2$ and S$^2$. Such kind of geometry has been previously found in the near horizon region of extremal accelerating Kerr black holes~\cite{Astorino:2016xiy}. The AdS$_2$ factor expressed in the first two terms of \eqref{nhmetric} is a universal part which also appears in the near horizon region of general extremal black holes~\cite{Kunduri:2013gce}. In studying the black hole physics near extremal horizons, AdS$_2$ plays essential roles. Many different sets of AdS$_2$ boundary conditions have been proposed for realizing different symmetries in the two dimensional Jackiw-Teitelboim gravity~\cite{Grumiller:2017qao}. Furthermore, new boundary conditions for AdS$_2$ have been proposed in~\cite{Godet:2020xpk} where the enhanced asymptotic symmetry group contains the local warped conformal symmetries, and the higher dimensional uplift of the new boundary conditions applied to extremal black holes also has been worked out in~\cite{Detournay:2023zni}. We will use the new boundary conditions to determine the asymptotic symmetry of the near horizon geometry \eqref{nhmetric}.

Following~\cite{Detournay:2023zni}, we use the Bondi-like coordinates
\be\label{Blct}
u=\tilde{t}+\frac{1}{\tilde{r}},~~~~\varphi=\frac{1}{b}\tilde{\phi}+\ln\tilde{r}\,,
\ee
in terms of which the near horizon metric takes the form
\be\label{nhumetric}
ds^2=\Gamma(x)\left(-\tilde{r}^2du^2-2dud\tilde{r}+\sigma^2(x)dx^2+\gamma^2(x)b^2(d\varphi+\tilde{r}du)^2\right)\,.
\ee
In this Bondi-like gauge, we consider the following boundary conditions
\be
\begin{pmatrix}
h_{uu}=\mathcal{O}(\tilde{r}) & h_{u\tilde{r}}=\mathcal{O}(\tilde{r}^{-1}) & h_{ux}=\mathcal{O}(\tilde{r}^{-1}) & h_{u\varphi}=\mathcal{O}(\tilde{r}^0)\\
h_{\tilde{r}u}=h_{u\tilde{r}} & h_{\tilde{r}\tilde{r}}=\mathcal{O}(\tilde{r}^{-3}) & h_{\tilde{r}x}=\mathcal{O}(\tilde{r}^{-2}) & h_{\tilde{r}\varphi}=\mathcal{O}(\tilde{r}^{-2})\\
h_{xu}=h_{ux} & h_{x\tilde{r}}=h_{\tilde{r}x} & h_{xx}=\mathcal{O}(\tilde{r}^{-1}) & h_{x\varphi}=\mathcal{O}(\tilde{r}^{-1})\\
h_{\varphi u}=h_{u\varphi} & h_{\varphi\tilde{r}}=h_{\tilde{r}\varphi} & h_{\varphi x}=h_{x\varphi} & h_{\varphi\varphi}=\mathcal{O}(\tilde{r}^{-1})
\end{pmatrix}\,,
\ee
where $h_{\mu\nu}$ is the metric perturbation which sets the specific radial falloff conditions at the asymptotic boundary $\tilde{r}\to\infty$. Due to the infinite scaling in \eqref{nhct}, the asymptotic boundary is still within the near horizon region. For these geometries, the asymptotic symmetries are specified by asymptotic Killing vectors that preserve the full metric up to the allowed perturbation
\be\label{aKvOeq}
\mathcal{L}_{\eta}(g_{\mu\nu}+h_{\mu\nu})=h_{\mu\nu}\,,
\ee
where $\mathcal{L}_{\eta}$ is the Lie derivative along $\eta$ and $g_{\mu\nu}$ is the near horizon metric \eqref{nhumetric}. The most general asymptotic Killing vector which satisfies \eqref{aKvOeq} takes the following form
\begin{align}\label{aKvO}
\eta=&\left(f(u)+\mathcal{O}(\tilde{r}^{-3})\right)\frac{\p}{\p u}+\left(-\tilde{r}f'(u)-g'(u)+\mathcal{O}(\tilde{r}^{-1})\right)\frac{\p}{\p\tilde{r}}\nn\\
&+\mathcal{O}(\tilde{r}^{-1})\frac{\p}{\p x}+\left(g(u)+\mathcal{O}(\tilde{r}^{-2})\right)\frac{\p}{\p\varphi}\,,
\end{align}
where $f(u)$ and $g(u)$ are arbitrary functions of $u$. The leading term in the asymptotic Killing vector will only perturb the AdS$_2$ factor in the original metric and the its result is consistent with the new boundary conditions discuss in~\cite{Godet:2020xpk}. Effectively, the leading asymptotic Killing vector keeps the form of the following class of metrics invariant
\be\label{nbc}
ds^2=\Gamma(x)\left((-\tilde{r}^2+2P(u)\tilde{r}+2T(u))du^2-2dud\tilde{r}+\sigma^2(x)dx^2+\gamma^2(x)b^2(d\varphi+\tilde{r}du)^2\right)\,,
\ee
where $P(u)$ and $T(u)$ are arbitrary functions of $u$ and transform under the Lie action of $\eta$ according to
\begin{align}
\delta_{\eta}P(u)&=f(u)P'(u)+f'(u)P(u)+f''(u)+g'(u)\,,\\
\delta_{\eta}T(u)&=f(u)T'(u)+2f'(u)T(u)-g'(u)P(u)+g''(u)\,.
\end{align}
As the AdS$_2$ boundary conditions proposed in~\cite{Godet:2020xpk} is related to the warped conformal symmetry, we will focus on the leading term in the asymptotic Killing vector which is
\be\label{aKv}
\eta=f(u)\frac{\p}{\p u}+(-\tilde{r}f'(u)-g'(u))\frac{\p}{\p\tilde{r}}+g(u)\frac{\p}{\p\varphi}\,,
\ee
and figure out its asymptotic symmetry on the background \eqref{nhumetric}. Correspondingly, the finite coordinate transformation that relates \eqref{nhumetric} and \eqref{nbc} can be written as
\be
u\to\mathcal{F}(u),~~~~\tilde{r}\to\frac{1}{\mathcal{F}'(u)}(\tilde{r}+\mathcal{G}'(u)),~~~~x\to x,~~~~\varphi\to\varphi-\mathcal{G}(u)\,,
\ee
where $\mathcal{F}(u)$ and $\mathcal{G}(u)$ are arbitrary functions of $u$, and they are related to $P(u)$ and $T(u)$ by
\begin{align}
P(u)&=-\mathcal{G}'(u)+\frac{\mathcal{F}''(u)}{\mathcal{F}'(u)}\,,\\
T(u)&=-\frac{1}{2}\mathcal{G}'(u)^2+\mathcal{G}'(u)\frac{\mathcal{F}''(u)}{\mathcal{F}'(u)}-\mathcal{G}''(u)\,.
\end{align}

The original Boyer-Lindquist coordinates $t$ and $\phi$ are defined on a torus which is determined by the identifications
\be
(t, \phi)\sim(t, \phi+2\pi)\sim(t+i\beta, \phi+i\beta\Omega_H)\,,\label{BLid}
\ee
where $\beta=1/T_H$ is the inverse horizon temperature. They are the spatial and thermal circles on which a finite temperature filed theory can be defined. By using coordinate transformations \eqref{nhct} and \eqref{Blct}, the Bondi-like coordinates $u$ and $\varphi$ will inherit the following identifications
\be
(u, \varphi)\sim\left(u, \varphi+\frac{2\pi}{b}\right)\sim\left(u+i\beta\frac{\delta}{r_0}, \varphi\right)\,.\label{tau}
\ee
In the extremal and near horizon limit, i.e., $\beta\to\infty$ and $\delta\to0$, the Bondi-like time coordinate $u$ can have finite periodic length which we denote it as $\tau$. If we Wick rotate to the Euclidean version of the near horizon geometry, the smoothness condition at the horizon requires $\tau$ to be $2\pi i$. In the later discussion, we will find this value is reasonable when compare the Kac-Moody level in the symmetry algebra to the non-extremal case. The leading asymptotic Killing vector \eqref{aKv} can be divided into two sets by taking the Fourier modes of the periodic functions $f(u)$ and $g(u)$
\begin{align}
f_n&=\eta\left(f(u)=\frac{\tau}{2\pi}e^{\frac{2\pi inu}{\tau}}, g(u)=0\right)\,,\label{fn}\\
g_n&=\eta\left(f(u)=0, g(u)=\frac{\tau}{2\pi i}e^{\frac{2\pi inu}{\tau}}\right)\,,\label{gn}
\end{align}
where $n$ is an integer. These two sets of mode vectors satisfy the following Virasoro and $U(1)$ Kac-Moody algebra without central extensions under Lie brackets
\begin{align}
i[f_m, f_n]&=(m-n)f_{m+n}\,,\nn\\
i[f_m, g_n]&=-ng_{m+n}\,,\label{LieB}\\
i[g_m, g_n]&=0\,.\nn
\end{align}

To figure out the possible central extensions, we will use the linearized covariant charges associated to the diffeomorphisms induced by the mode vectors \eqref{fn} and \eqref{gn}. The charges are acting on the asymptotic boundary and implement the asymptotic symmetry algebra through Dirac brackets. On a background geometry $g$, the variation of covariant charge associated to a vector field $\zeta$ in general relativity can be expressed as
\be\label{dQ}
\delta\mathcal{Q}(\zeta, h; g)=\frac{1}{16\pi}\int_{\p\Sigma}*F_{IW}\,,
\ee
where the two-form field $F_{IW}$ has components given by~\cite{Iyer:1994ys}
\be\label{FIW}
(F_{IW})_{\mu\nu}=\frac{1}{2}\nabla_{\mu}\zeta_{\nu}h+\nabla_{\mu}h^{\sigma}_{~\nu}\zeta_{\sigma}+\nabla_{\sigma}\zeta_{\mu}h^{\sigma}_{~\nu}+\nabla_{\sigma}h^{\sigma}_{~\mu}\zeta_{\nu}-\nabla_{\mu}h\zeta_{\nu}-(\mu\leftrightarrow\nu)\,,
\ee
$h^{\mu\nu}$ is the variation of inverse metric $h^{\mu\nu}=\delta g^{\mu\nu}$ and $h=g_{\mu\nu}h^{\mu\nu}$. The co-dimension two surface of integration $\p\Sigma$ is chosen as the asymptotic boundary where $\tilde{r}\to\infty$ and $\varphi$ is kept as a constant. The integration is then performed along $u$ and $x$ in finite regions $(0, \tau)$ and $(-1, 1)$. The dual warped CFT in this extremal case is defined at the asymptotic boundary of the near horizon geometry with coordinates $\varphi$ and $u$. In a warped CFT, the time direction usually takes along the $U(1)$ direction~\cite{Detournay:2012pc}. In the current set up, as indicating in the asymptotic Killing vector \eqref{aKv}, the warped CFT $U(1)$ symmetry is generated by $\p/\p\varphi$ and the $SL(2 R)$ conformal symmetry is related to $\p/\p u$. So from the warped CFT perspective, the constant $\varphi$ slice is a constant warped CFT time slice and the charges are integrated over the warped CFT spatial coordinate $u$ with periodicity from the horizon smoothness condition. The bulk coordinate $x$ is invisible from the boundary theory which should be averaged by a full integration.

The variation of the charge \eqref{dQ} is integrable with the boundary conditions satisfied by the asymptotic Killing vector \eqref{aKv}. The variation of the inverse metric satisfying the same boundary conditions can be written as
\be
h^{\mu\nu}=\frac{\delta g^{\mu\nu}}{\delta P(u)}\delta P(u)+\frac{\delta g^{\mu\nu}}{\delta T(u)}\delta T(u)\,,
\ee
where $g^{\mu\nu}$ is the inverse metric of \eqref{nbc}. Then the corresponding variation of the charge takes the form
\be
\delta Q(\eta, h; g)=\frac{1}{8\pi}\int_{-1}^1\int_0^{\tau}b\Gamma(x)\sigma(x)\gamma(x)\left(g(u)\delta P(u)-f(u)\delta T(u)\right)dxdu\,,
\ee
and the above variation can be integrated to obtain the finite charge
\be
Q(\eta, h; g)=\frac{1}{8\pi}\int_{-1}^1\int_0^{\tau}b\Gamma(x)\sigma(x)\gamma(x)\left(g(u)P(u)-f(u)T(u)\right)dxdu\,.
\ee
For any two vector fields $\xi$ and $\zeta$, given the charge integrability, the covariant charges form an algebra under Dirac bracket~\cite{Compere:2018aar}
\be\label{DB}
\{\mathcal{Q}_{\xi}, \mathcal{Q}_{\zeta}\}=\mathcal{Q}_{[\xi, \zeta]}+K_{\zeta, \xi},~~~~K_{\zeta, \xi}=\delta\mathcal{Q}(\xi, \mathcal{L}_{\zeta}g; g)\,,
\ee
where ``$[~,~]$'' is the Lie bracket, $K_{\zeta, \xi}$ is the possible central extension term, and $\mathcal{L}_{\zeta}$ is the Lie derivative along $\zeta$. Let us define the charge variation associated to the mode vectors \eqref{fn} and \eqref{gn} as
\be\label{dLndPnext}
\delta L^{ext}_n=\delta\mathcal{Q}(f_n, h; g),~~~~\delta P^{ext}_n=\delta\mathcal{Q}(g_n, h; g)\,.
\ee
These charges are integrable and the corresponding charge algebra in the form of \eqref{DB} can be divided into Virasoro, mixed, and Kac-Moody sectors with possible central extensions
\begin{align}
\{L^{ext}_n, L^{ext}_m\}&=(m-n)L^{ext}_{m+n}+K^{ext}_{m, n},~~~~K^{ext}_{m, n}=\delta\mathcal{Q}(f_n, \mathcal{L}_{f_m}g; g)\,,\nn\\
\{L^{ext}_n, P^{ext}_m\}&=mP^{ext}_{m+n}+\mathfrak{R}^{ext}_{m, n},~~~~\mathfrak{R}^{ext}_{m, n}=\delta{Q}(f_n, \mathcal{L}_{g_m}g; g)\,,\label{DiracB}\\
\{P^{ext}_n, P^{ext}_m\}&=k^{ext}_{m, n},~~~~k^{ext}_{m, n}=\delta\mathcal{Q}(g_n, \mathcal{L}_{g_m}g; g)\,.\nn
\end{align}
Performing the integration \eqref{dQ} at the asymptotic boundary of the background geometry \eqref{nhumetric} with mode vectors \eqref{fn} and \eqref{gn}, the central extension terms in the above charge algebra can be calculated
\begin{align}
\delta\mathcal{Q}(f_n, \mathcal{L}_{f_m}g; g)&=0\,,\\
\delta{Q}(f_n, \mathcal{L}_{g_m}g; g)&=\frac{1}{16\pi}\int_{-1}^1\int_0^{\tau}-2im^2b\Gamma(x)\sigma(x)\gamma(x)e^{\frac{2\pi i(m+n)u}{\tau}}dxdu\nn\\
&=i\frac{\lambda^2r_+a\tau}{2\pi k_+(\lambda-r_+\alpha(a+l))(\lambda+r_+\alpha(a-l))}m^2\delta_{m, -n}\,,\\
\delta\mathcal{Q}(g_n, \mathcal{L}_{g_m}g; g)&=\frac{1}{16\pi}\int_{-1}^1\int_0^{\tau}-\frac{i\tau m}{\pi}b\Gamma(x)\sigma(x)\gamma(x)e^{\frac{2\pi i(m+n)u}{\tau}}dxdu\nn\\
&=i\frac{\lambda^2r_+a\tau^2}{4\pi^2k_+(\lambda-r_+\alpha(a+l))(\lambda+r_+\alpha(a-l))}m\delta_{m, -n}\,.
\end{align}
The charge algebra for the generators defined in \eqref{dLndPnext} thus can be expressed in a concise form
\begin{align}
\{L^{ext}_n, L^{ext}_m\}&=(m-n)L^{ext}_{m+n}+i\frac{c^{ext}}{12}m^3\delta_{m, -n}\,,\nn\\
\{L^{ext}_n, P^{ext}_m\}&=mP^{ext}_{m+n}+i\varkappa^{ext}m^2\delta_{m, -n}\,,\label{DiracBcc}\\
\{P^{ext}_n, P^{ext}_m\}&=i\frac{\kappa^{ext}}{2}m\delta_{m, -n}\,,\nn
\end{align}
This is the warped conformal algebra consists of a Virasoro and an $U(1)$ Kac-Moody algebra with the Virasoro central charge $c^{ext}$, the mixed central charge $\varkappa^{ext}$, and the Kac-Moody level $\kappa^{ext}$ given by
\begin{align}
c^{ext}&=0\,,\\
\varkappa^{ext}&=\frac{\lambda^2r_+a\tau}{2\pi k_+(\lambda-r_+\alpha(a+l))(\lambda+r_+\alpha(a-l))}\,,\\
\kappa^{ext}&=\frac{\lambda^2r_+a\tau^2}{2\pi^2k_+(\lambda-r_+\alpha(a+l))(\lambda+r_+\alpha(a-l))}\,.
\end{align}
The mixed central extension term in \eqref{DiracBcc} however can be removed by the following charge redefinition~\cite{Godet:2020xpk}
\be
L^{*ext}_n=L^{ext}_n+\frac{2\varkappa^{ext}}{\kappa^{ext}}nP^{ext}_n,~~~~P^{*ext}_n=P^{ext}_n\,.
\ee
under which the new charge algebra takes the form
\begin{align}
\{L^{*ext}_n, L^{*ext}_m\}&=(m-n)L^{*ext}_{m+n}+i\frac{c^{*ext}}{12}m^3\delta_{m, -n}\,,\nn\\
\{L^{*ext}_n, P^{*ext}_m\}&=mP^{*ext}_{m+n}\,,\label{DiracBccs}\\
\{P^{*ext}_n, P^{*ext}_m\}&=i\frac{\kappa^{*ext}}{2}m\delta_{m, -n}\,.\nn
\end{align}
This is the standard warped conformal algebra with the Virasoro central charge $c^{*ext}$ and the Kac-Moody level $\kappa^{*ext}$ given by
\begin{align}
c^{*ext}&=\frac{12\lambda^2r_+a}{k_+(\lambda-r_+\alpha(a+l))(\lambda+r_+\alpha(a-l))}\,,\label{cext}\\
\kappa^{*ext}&=\frac{\lambda^2r_+a\tau^2}{2\pi^2k_+(\lambda-r_+\alpha(a+l))(\lambda+r_+\alpha(a-l))}\,.\label{kappaext}
\end{align}

\subsection{Non-extremal case}
In the non-extremal case, the temperature \eqref{TH} does not vanish and the horizon radius $r_+$ is a single zero of the horizon polynomial $Q$. In contrast to the extremal case, the near horizon geometry is approximated by a Rindler spacetime in the non-extremal case instead of the one with an AdS$_2$ fiber. The Rindler spacetime contains no conformal symmetry, however, the dynamics of near horizon low frequency scalar perturbations on the non-extremal background \eqref{GPDmetric} is governed by the hidden conformal symmetry \eqref{Casimireq}. The global warped conformal symmetry can be generated by $H_{0, \pm}$ and $\bar{H}_0$ defined in \eqref{Hp0m} and \eqref{bHp0m}. As discussed in~\cite{Xu:2023tfl, Aggarwal:2019iay}, for non-extremal black holes, the infinite local symmetries can be recovered from the diffeomorphisms that keep the radial equation in the near horizon region as well as the frequency of the scalar perturbation invariant. We will follow this prescription to figure out the local warped conformal symmetries for the non-extremal background \eqref{GPDmetric}

Consider a set of diffeomorphisms induced by the following vector fields in conformal coordinates
\begin{align}
\xi(l(\omega^+))&=l(\omega^+)\frac{\p}{\p\omega^+}+\frac{\p l(\omega^+)}{\p\omega^+}\frac{y}{2}\frac{\p}{\p y}\,,\label{xi}\\
\zeta(p(\omega^+))&=p(\omega^+)\left(\omega^-\frac{\p}{\p\omega^-}+\frac{y}{2}\frac{\p}{\p y}\right)\,,\label{zeta}
\end{align}
where $l(\omega^+)$ and $p(\omega^+)$ are chosen functions such that the above vector fields are periodic under the identifications \eqref{ccids}. This is equivalent to require
\be\label{lppc}
l(e^{4\pi^2T_R}\omega^+)=e^{4\pi^2T_R}l(\omega^+)\,,~~~~p(e^{4\pi^2T_R}\omega^+)=p(\omega^+)\,.
\ee
The form of the vector fields \eqref{xi} and \eqref{zeta} are chosen so that the induced infinitesimal coordinate transformations can keep the $SL(2, R)$ quadratic Casimir $\mathcal{H}^2$ \eqref{Casimir} and the $U(1)$ generator $\bar{H}_0$ \eqref{bHp0m} invariant. This means that the vector fields keep the scalar radial equation in the near horizon region \eqref{Casimireq} as well as the eigenvalue of $\bar{H}_0$ which is proportional to the scalar frequency invariant. These vector fields which preserve the dynamics of the scalar field can be understood as the non-extremal analog of the asymptotic Killing vector \eqref{aKv} which preserve the near horizon geometry. Invoking the Fourier modes derived from the periodic condition \eqref{lppc}, one can define the mode vectors
\begin{align}
l_n&=\xi\left(l(\omega^+)=2\pi T_R(\omega^+)^{1+\frac{in}{2\pi T_R}}\right)\,,\label{ln}\\
p_n&=\zeta\left(p(\omega^+)=(\omega^+)^{\frac{in}{2\pi T_R}}\right)\,,\label{pn}
\end{align}
where $n$ is an integer. These mode vectors also satisfy the classical Virasoro and $U(1)$ Kac-Moody algebra
\begin{align}
i[l_m, l_n]&=(m-n)l_{m+n}\,,\nn\\
i[l_m, p_n]&=-np_{m+n}\,,\label{LieBnext}\\
i[p_m, p_n]&=0\,.\nn
\end{align}

The possible central extensions in this case can be carried out by the method analyzing the phase space of the vector fields \eqref{xi} and \eqref{zeta} in the near horizon region~\cite{Aggarwal:2019iay}, which is initiated form the calculation for the central charges of two copies of the Virasoro algebra on an non-extremal Kerr background~\cite{Haco:2018ske}. The linearized covariant charges associated to the diffeomorphisms induced by the vector fields \eqref{xi} and \eqref{zeta} are now evaluated at the horizon bifurcation surface. The variation of the covariant charge contains two parts
\be\label{IWpct}
\delta\mathcal{Q}=\delta\mathcal{Q}_{IW}+\delta\mathcal{Q}_{ct}\,.
\ee
The first term is the Iyer-Wald charge~\cite{Iyer:1994ys} with its explicit form expressed in \eqref{dQ} and \eqref{FIW}. The surface of integration $\p\Sigma$ is chosen as the horizon bifurcation surface, which is the intersection of the future and past horizons. In the Boyer-Lindquist coordinates, the future horizon is located at $r=r_+, t\in(0, \infty)$ and the past horizon is located at $r=r_+, t\in(-\infty, 0)$. In terms of the conformal coordinates defined in \eqref{cc}, these correspond to $\omega^-=0$ and $\omega^+=0$, respectively. So the horizon bifurcation is located at $\omega^+=\omega^-=0$. The geometry near this surface can be expressed as a double expansion of the original metric \eqref{GPDmetric} around $\omega^+=\omega^-=0$ in conformal coordinates
\begin{align}
ds^2=&\frac{F_{+-}}{y^2}d\omega^+d\omega^s+\frac{F_{yy}}{y^2}dy^2+F_{xx}dx^2+\omega^+\frac{F_{-y}}{y^3}d\omega^-dy+\omega^-\frac{F_{+y}}{y^3}d\omega^+dy\nn\\
&+\mathcal{O}((\omega^+)^2, \omega^+\omega^-, (\omega^-)^2)\,,\label{nbifg}
\end{align}
where
\begin{align}
F_{+-}&=\frac{4\lambda^2\left(r_+^2+(l-ax)^2\right)}{k_+\left(\lambda-r_+\alpha(l-ax)\right)^2}\,,\nn\\
F_{yy}&=\frac{4\lambda^2(r_++r_s)^2a^2(1-x^2)P}{k_+^2\left(r_+^2+(l-ax)^2\right)\left(\lambda-r_+\alpha(l-ax)\right)^2}\,,\nn\\
F_{xx}&=\frac{\lambda^2\left(r_+^2+(l-ax)^2\right)}{\left(\lambda-r_+\alpha(l-ax)\right)^2(1-x^2)P}\,,\label{FF}\\
F_{-y}&=\frac{-4\lambda^2(r_+^2-r_s^2)\left[k_+\left(r_+^2+(l-ax)^2\right)+a^2(1-x^2)P\right]}{k_+^2\left(r_+^2+(l-ax)^2\right)\left(\lambda-r_+\alpha(l-ax)\right)^2}\,,\nn\\
F_{+y}&=\frac{-4\lambda^2\left[k_+\left(r_+^2+r_s^2+2(l-ax)^2\right)\left(r_+^2+(l-ax)^2\right)-(3r_+^2+4r_+r_s+r_s^2)a^2(1-x^2)P\right]}{k_+^2\left(r_+^2+(l-ax)^2\right)\left(\lambda-r_+\alpha(l-ax)\right)^2}\,.\nn
\end{align}
The second term in \eqref{IWpct} is a counterterm which is introduced for some consistency conditions like charge associativity and central extension terms in the charge algebra being state independent. The specific form of the counterterm here is further demanded that it can eliminate the mixed central extension term to realize a standard warped conformal algebra like \eqref{DiracBccs} in the phase space. The counterterm has different forms in different contexts. For example, it is specified as the Wald-Zoupas counterterm for Kerr black holes in deriving the two copies of the Virasoro algebra~\cite{Haco:2018ske, Haco:2019ggi} and the standard warped conformal algebra~\cite{Xu:2023tfl, Aggarwal:2019iay}. For Kerr-AdS or Kerr-NUT black holes, a more general counterterm is required in getting the Virasoro central charges~\cite{Perry:2020ndy, Perry:2022udk}. In this paper, for the black hole-like spacetime \eqref{GPDmetric}, we propose another counterterm to derive the standard warped conformal algebra in the phase space of the vector fields \eqref{xi} and \eqref{zeta}. The form of the counterterm associated to a vector field $\xi$ can be expressed as
\be\label{ct}
\delta\mathcal{Q}_{ct}(\xi, h; g)=\frac{1}{16\pi}\int_{\p\Sigma}i_{\xi}\cdot(~^*X)\,,
\ee
where $X$ is a spacetime one-form defined by
\be
X=2h^{\nu}_{~\mu}\nabla_{\rho}N^{\rho}_{~\nu}dx^{\mu}\,,
\ee
$N_{\mu\nu}$ are components of the volume two-form $N$ on the normal bundle of the integration surface $\p\Sigma$. $X$ is linear in the inverse metric variation $h^{\mu\nu}$ which indicates that it might be raised from the ambiguity in defining a symplectic potential~\cite{Wald:1999wa}. Taking $\p\Sigma$ as the horizon bifurcation surface, the volume form on its normal bundle is given by
\be
N=N_{\mu\nu}dx^{\mu}\wedge dx^{\nu}=-(l_{\mu}n_{\nu}-n_{\mu}l_{\nu})|_{\p\Sigma}dx^{\mu}\wedge dx^{\nu}\,,
\ee
where
\be\label{lmunmu}
l_{\mu}|_{\p\Sigma}dx^{\mu}=y^{\frac{-2T_L}{T_R+T_L}}d\omega^-,~~~~n_{\mu}|_{\p\Sigma}dx^{\mu}=-\frac{F_{+-}}{2}y^{\frac{-2T_R}{T_R+T_L}}d\omega^+\,,
\ee
are the null normal co-vectors of the future and past horizons, respectively, and satisfy $l^{\mu}n_{\mu}|_{\p\Sigma}=-1$. The $y$ dependent factors in the definitions of the co-vectors are set for the periodicity under identifications \eqref{ccids}. With the metric expansion \eqref{nbifg} and the normal co-vectors \eqref{lmunmu}, one can verify that
\be
N^{\mu}_{~\nu}=
\begin{pmatrix}
~1~&~0~&~0~&~0~\\
~0~&~-1~&~0~&~0~\\
~0~&~0~&~0~&~0~\\
~0~&~0~&~0~&~0~
\end{pmatrix}\,.
\ee

Let us define the charge variation with counterterm associated to the mode vectors \eqref{ln} and \eqref{pn} as
\be
\delta L_n=\delta\mathcal{Q}(l_n, h; g),~~~~\delta P_n=\delta\mathcal{Q}(p_n, h; g)\,.
\ee
In the form of \eqref{DB}, the algebra for these charges also includes the Virasoro, mixed, and Kac-Moody sectors with possible central extensions
\begin{align}
\{L_n, L_m\}&=(m-n)L_{m+n}+K_{m, n},~~~~K_{m, n}=\delta\mathcal{Q}(l_n, \mathcal{L}_{l_m}g; g)\,,\nn\\
\{L_n, P_m\}&=mP_{m+n}+\mathfrak{R}_{m, n},~~~~\mathfrak{R}_{m, n}=\delta{Q}(l_n, \mathcal{L}_{p_m}g; g)\,,\label{DiracBnext}\\
\{P_n, P_m\}&=k_{m, n},~~~~k_{m, n}=\delta\mathcal{Q}(p_n, \mathcal{L}_{p_m}g; g)\,.\nn
\end{align}
Using the metric expansion \eqref{nbifg} and the mode vectors \eqref{ln} and \eqref{pn}, one can evaluate the charge variation defined in \eqref{IWpct} with the Iyer-Wald charge \eqref{dQ} and the counterterm \eqref{ct} to carry out the central extension terms in the above algebra. In calculating the charge variation, one need to perform integrations on the horizon bifurcation surface. These integrations will only receive non-vanishing contributions from the simple poles at $\omega^+$. For a reference point $\omega^+_0$ near $\omega^+$, the relevant part of the contribution has the form
\be
\int_{\omega_0^+}^{e^{4\pi^2T_R}\omega_0^+}(\omega^+)^{-1+\frac{i(m+n)}{2\pi T_R}}d\omega^+=4\pi^2T_R\delta_{m, -n}\,.
\ee
Taking this into account, the central term in the Virasoro sector in \eqref{DiracBnext} contains the following two parts
\begin{align}
\delta\mathcal{Q}_{IW}(l_n, \mathcal{L}_{l_m}g; g)&=i\left((4\pi^2T_R^2)m+m^3\right)\delta_{m, -n}\int_{-1}^1dx\frac{\sqrt{F_{yy}F_{xx}}(2F_{+-}+F_{-y}-F_{+y})}{16F_{+-}}\,,\\
\delta\mathcal{Q}_{ct}(l_n, \mathcal{L}_{l_m}g; g)&=-i\left((4\pi^2T_R^2)m+m^3\right)\delta_{m, -n}\int_{-1}^1dx\frac{\sqrt{F_{yy}F_{xx}}(F_{-y}-F_{+y})}{16F_{+-}}\,.\label{Vct}
\end{align}
Adding these two parts together and absorbing the linear term in $m$ into $L_0$, we get the central extension term for the Virasoro sector
\begin{align}
K_{m, n}&=\frac{i}{8}m^3\delta_{m, -n}\int_{-1}^1dx\sqrt{F_{yy}F_{xx}}\nn\\
&=i\frac{\lambda^2(r_++r_s)a}{2k_+\left(\lambda-r_+\alpha(a+l)\right)\left(\lambda+r_+\alpha(a-l)\right)}m^3\delta_{m, -n}\,.
\end{align}
As one can see that the counterterm is not unique since it always takes an integration along $x$. The integrand in counterterm may has additional nonzero terms which have zero $x$ integration. The Wald-Zoupas counterterm~\cite{Haco:2018ske} takes a similar form as in \eqref{ct}, but with the spacetime one-form $X$ given by
\be
X_{WZ}=2h^{\nu}_{~\mu}q^{\rho}_{~\nu}n^{\sigma}\nabla_{\rho}l_{\sigma}dx^{\mu}\,,
\ee
where $q_{\mu\nu}=g_{\mu\nu}+l_{\mu}n_{\nu}+n_{\mu}l_{\nu}$ is the induced metric on $\p\Sigma$. By the same token, the Wald-Zoupas counterterm for the Virasoro sector also can be calculated with the following form
\begin{align}
\delta\mathcal{Q}_{WZ}(l_n, \mathcal{L}_{l_m}g; g)&=-i\left((4\pi^2T_R^2)m+m^3\right)\delta_{m, -n}\nn\\
&\times\int_{-1}^1dx\frac{\sqrt{F_{yy}F_{xx}}\left(2(T_R-T_L)F_{+-}+(T_R+T_L)(F_{-y}-F_{+y})\right)}{32(T_R+T_L)F_{+-}}\,.\label{VWZ}
\end{align}
In the Kerr case where $\lambda=a$, $e=g=\alpha=l=\Lambda=0$, and $r_s=r_-$, one can easily show that the Wald-Zoupas counterterm \eqref{VWZ} gives exactly the result in \eqref{Vct} after the $x$ integration. Similar for the other two sectors. The central term in the mixed sector in \eqref{DiracBnext} has the following two parts with opposite signs
\begin{align}
\delta\mathcal{Q}_{IW}(l_n, \mathcal{L}_{p_m}g; g)&=\left((2\pi iT_R)m+m^2\right)\delta_{m, -n}\int_{-1}^1dx\frac{\sqrt{F_{yy}F_{xx}}(F_{-y}-F_{+y})}{16F_{+-}}\,,\\
\delta\mathcal{Q}_{ct}(l_n, \mathcal{L}_{p_m}g; g)&=-\left((2\pi iT_R)m+m^2\right)\delta_{m, -n}\int_{-1}^1dx\frac{\sqrt{F_{yy}F_{xx}}(F_{-y}-F_{+y})}{16F_{+-}}\,.
\end{align}
These two parts cancel with each other and give vanishing central extension term for the mixed sector
\be
\mathfrak{R}_{m, n}=0\,.
\ee
The cancellation of the mixed central extension term does not depend on the details of the functions \eqref{FF}, so the counterterm \eqref{ct} we proposed here might work for a large class of spacetime geometry that endowed with near horizon bifurcation surface expansions in the form of \eqref{nbifg}. The central term in the Kac-Moody sector in \eqref{DiracBnext} has the following two terms
\begin{align}
\delta\mathcal{Q}_{IW}(p_n, \mathcal{L}_{p_m}g; g)&=-im\delta_{m, -n}\int_{-1}^1dx\frac{\sqrt{F_{yy}F_{xx}}(2F_{+-}-F_{-y}+F_{+y})}{16F_{+-}}\,,\\
\delta\mathcal{Q}_{ct}(p_n, \mathcal{L}_{p_m}g; g)&=-im\delta_{m, -n}\int_{-1}^1dx\frac{\sqrt{F_{yy}F_{xx}}(F_{-y}-F_{+y})}{16F_{+-}}\,.
\end{align}
Sum of the above two terms gives the central extension term for the Kac-Moody sector
\begin{align}
k_{m, n}&=-\frac{i}{8}m\delta_{m, -n}\int_{-1}^1dx\sqrt{F_{yy}F_{xx}}\nn\\
&=-i\frac{\lambda^2(r_++r_s)a}{2k_+\left(\lambda-r_+\alpha(a+l)\right)\left(\lambda+r_+\alpha(a-l)\right)}m\delta_{m, -n}\,.
\end{align}
In all, the charge algebra induced from the commutation relations \eqref{LieBnext} take the form as the standard warped conformal algebra
\begin{align}
\{L_n, L_m\}&=(m-n)L_{m+n}+i\frac{c}{12}m^3\delta_{m, -n}\,,\nn\\
\{L_n, P_m\}&=mP_{m+n}\,,\label{DiracBccsnext}\\
\{P_n, P_m\}&=i\frac{\kappa}{2}m\delta_{m, -n}\,,\nn
\end{align}
with the Virasoro central charge $c$ and the Kac-Moody level $\kappa$ given by
\begin{align}
c&=\frac{6\lambda^2(r_++r_s)a}{k_+\left(\lambda-r_+\alpha(a+l)\right)\left(\lambda+r_+\alpha(a-l)\right)}\,,\label{cnext}\\
\kappa&=-\frac{\lambda^2(r_++r_s)a}{k_+\left(\lambda-r_+\alpha(a+l)\right)\left(\lambda+r_+\alpha(a-l)\right)}\,.\label{kappanext}
\end{align}
Taking the extremal limit where $r_+=r_s$, one find that the non-extremal central charge $c$ \eqref{cnext} reproduce exactly the extremal central charge $c^{*ext}$ \eqref{cext}, while the non-extremal Kac-Moody level $\kappa$ \eqref{kappanext} will match the extremal Kac-Moody level $\kappa^{*ext}$ \eqref{kappaext} in the case that
\be
\tau=2\pi i\,.
\ee
As claimed below \eqref{tau}, the Bondi-like time coordinate $u$ can have finite periodicity $\tau$ which now can be fixed as $2\pi i$ for matching the extremal and non-extremal central extension terms in the warped conformal algebras derived with different methods.

\section{Horizon entropy from warped conformal symmetries}\label{sec4}
Either in the extremal or in the non-extremal case, the horizon has non-vanishing area and thus has non-vanishing entropy. In this section, we will use the entropy formula of a finite temperature warped CFT derived from the modular properties of its partition function to recover the entropy of the horizon at $r=r_+$ as expressed in \eqref{SH}. The coordinates $t^+$ and $t^-$ of the warped CFT are defined as
\begin{align}
t^+=2\pi T_R\phi+2n_Rt,~~~~t^-=-2\pi T_L\phi-2n_Lt\,.
\end{align}
It is a natural choice of the field theory coordinates. As one can verify that
\be
e^{-i\omega t+im\phi}=e^{-i\omega_Rt^+-i\omega_Lt^-}\,,
\ee
where
\be
\omega_R=\alpha_++\alpha_s,~~~~\omega_L=\alpha_+-\alpha_s\,,
\ee
with $\alpha_+$ and $\alpha_s$ being the monodromies around the branch cuts of the radial solutions \eqref{mono}, the $U(1)\times U(1)$ part of the global warped conformal symmetry generated by
\be
H_0=i\frac{\p}{\p t^+},~~~~\bar{H}_0=-i\frac{\p}{\p t^-}\,,
\ee
becomes the translational symmetry along these two coordinates with momentum eigenvalues $\omega_R$ and $\omega_L$, respectively. The phase factor in the scalar field by the $2\pi$ identification along $\phi$ can be expressed as
\be
e^{i2\pi m}=e^{-i4\pi^2T_R\omega_R+i4\pi^2T_L\omega_L}\,,
\ee
from which the parameters $T_R$ and $T_L$ given in \eqref{TLTR} can be recognized as the thermal parameters in the dual field theory with coordinates $t^+$ and $t^-$.

Using relations \eqref{TH}, \eqref{TLTR}, and \eqref{OH}, the spatial and thermal circles defined in the original Boyer-Lindquist coordinates \eqref{BLid} can be translated into those in the coordinates $t^+$ and $t^-$
\be\label{tptmid}
(t^+, t^-)\sim(t^++4\pi^2T_R, t^--4\pi^2T_L)\sim(t^++2\pi i, t^-+2\pi i)\,.
\ee
The above identifications define a generic torus where the warped CFT lives on. By further performing a warped conformal transformation
\be\label{htphtm}
\hat{t}^+=-\frac{t^+}{2\pi T_R}\,,~~~~\hat{t}^-=t^-+\frac{T_L}{T_R}t^+\,,
\ee
the generic torus \eqref{tptmid} can be transformed into a canonical one in terms of the new coordinates $\hat{t}^+$ and $\hat{t}^-$
\be\label{htphtmid}
(\hat{t}^+, \hat{t}^-)\sim(\hat{t}^+-2\pi, \hat{t}^-)\sim(\hat{t}^+-2\pi\hat{\tau}, \hat{t}^-+2\pi\hat{\bar{\tau}})\,,
\ee
where
\be\label{hthtb}
\hat{\tau}=\frac{i}{2\pi T_R}\,,~~~~\hat{\bar{\tau}}=\frac{i(T_R+T_L)}{T_R}\,.
\ee
On the canonical torus \eqref{htphtmid}, the thermal entropy of the warped CFT takes the form~\cite{Detournay:2012pc}
\be\label{S01htbht}
S_{(0|1)}(\hat{\bar{\tau}}|\hat{\tau})=2\pi i\frac{\hat{\bar{\tau}}}{\hat{\tau}}\hat{P}_0^{vac}+4\pi i\frac{1}{\hat{\tau}}\hat{L}_0^{vac}\,,
\ee
where $\hat{L}_0^{vac}$ and $\hat{P}_0^{vac}$ are the vacuum expectation values of the Virasoro and Kac-Moody zero modes on the canonical torus, respectively. These two vacuum values are somehow related through a spectral flow parameter which labels the vacuum state of a warped CFT. To see this, it is convenient to map the canonical torus back to a Lorentzian plane, where the vacuum values of zero modes are zeros, with the following warped conformal transformation
\be\label{Lplane}
\bar{t}^+=e^{i\hat{t}^+}\,,~~~~\bar{t}^-=\hat{t}^-+2\mu\hat{t}^+\,,
\ee
where $\mu$ is the spectral flow parameter. The Virasoro and Kac-Moody modes ($\hat{L}_n, \hat{P}_n$) on the canonical torus and ($\bar{L}_n, \bar{P}_n$) on the plane are related under the above transformation~\cite{Detournay:2012pc}
\begin{align}
\hat{L}_n&=\bar{L}_n+2\mu\bar{P}_n+\left(\kappa\mu^2-\frac{c}{24}\right)\delta_{n, 0}\\
\hat{P}_n&=\bar{P}_n+\kappa\mu\delta_{n, 0}\,,
\end{align}
where $c$ is the Virasoro central charge and $\kappa$ is the Kac-Moody level. Given that zero modes have vanishing vacuum expectation values defined on the plane, i.e. $\bar{L}_0^{vac}=\bar{P}_0^{vac}=0$, the zero modes on the canonical torus take the following vacuum values
\be\label{LPvacrela}
\hat{L}_0^{vac}=\kappa\mu^2-\frac{c}{24}\,,~~~~\hat{P}_0^{vac}=\kappa\mu\,.
\ee
The vacuum state labeled by different $\mu$ gives different vacuum values of the zero modes. Now, in order to recover the bulk horizon entropy, we choose a specific vacuum state of the warped CFT with 
\be\label{mu0}
\mu^2=\frac{c}{24\kappa}\,,
\ee
and this implies $\hat{L}_0^{vac}=0$. Since the Virasoro zero mode on the canonical torus is dual to the conserved charge associated to $\p/\p\hat{t}^+=-\p/\p\phi$ which is proportional to the angular momentum of the background \eqref{GPDmetric}, such a vacuum state is dual to a bulk spacetime with zero angular momentum. With this vacuum state, and substituting the results of $c$ \eqref{cnext} and $\kappa$ \eqref{kappanext} derived from the non-extremal case into \eqref{LPvacrela}, one can write down the following vacuum values of the zero modes
\be
\hat{L}_0^{vac}=0,~~~~(\hat{P}_0^{vac})^2=-\left(\frac{\lambda^2(r_++r_s)a}{2k_+\left(\lambda-r_+\alpha(a+l)\right)\left(\lambda+r_+\alpha(a-l)\right)}\right)^2\,.
\ee
With the above vacuum values and the thermal parameters \eqref{hthtb} and \eqref{TLTR}, the thermal entropy of the warped CFT given in \eqref{S01htbht} can be evaluated as
\be
S_{(0|1)}(\hat{\bar{\tau}}|\hat{\tau})=4\pi^2|\hat{P}_0^{vac}|(T_R+T_L)=\frac{\pi\lambda^2(r_+^2+(a+l)^2)}{(\lambda-r_+\alpha(a+l))(\lambda+r_+\alpha(a-l))}\,,
\ee
which exactly agrees with the horizon entropy calculated by the area law \eqref{SH}.

\section{Summary and Discussion}\label{sec5}
In this paper, we analyze the underlying symmetry of the family of black hole-like spacetimes \eqref{GPDmetric} generalized from the Pleba\'nski-Demia\'nski solution. Such kind of a spacetime is labeled by seven physical parameters which characterize the black hole mass, electric and magnetic charges, angular momentum, acceleration, NUT charge, and cosmological constant. Physically, this configuration describes two black holes and the metric contains conical singularities at the poles of its axis. The corresponding deficit or excess angles at the poles can be understood as the effects induced by two cosmic strings linking the two black holes to infinity or a strut connecting them. These effective objects can be viewed as the cause of the constant relative acceleration between the two black holes.

The horizon polynomial of the metric which indicates the location of the horizons by its zeros is a quartic function of the radius. We focus on one of its zeros and discuss the hidden conformal symmetry near the corresponding horizon. The hidden conformal symmetry can be carried out by considering the wave equation of a perturbative massless scalar field. The scalar field is conformally coupled to the gravity such that after a proper conformal transformation its transformed wave equation become separable. The separated radial and angular equations are all Heun-type equations. In the near horizon region, the radial equation can be approximated by a hypergeometric equation when the scalar frequency is small which indicates there must be conformal symmetry in controlling the dynamics of the scalar field. By invoking the conformal coordinates, the low frequency scalar radial equation in the near horizon region can be further casted into the form of a Casimir equation of $SL(2, R)$ with two copies. The hidden conformal symmetry $SL(2, R)\times SL(2, R)$ has a subgroup $SL(2, R)\times U(1)$ that keep the radial equation as well as the scalar frequency invariant. This part of the symmetry is viewed as the global warped conformal symmetry of a warped CFT. The global symmetry is spontaneously broken to $U(1)\times U(1)$ due to the $2\pi$ periodicity along the coordinate $\phi$ which in turn plays the role as the translational symmetry along the two coordinates defined in the warped CFT.

The local warped conformal symmetries can be obtained by analyzing the phase space of specific vector fields defined in the asymptotic region of the near horizon extremal geometry \eqref{nhumetric} or in the near horizon region of the non-extremal background \eqref{GPDmetric}. There are two different methods to carry out the local symmetries. In the extremal case, the near horizon region can be mapped into an infinite spacetime which has an AdS$_2$ factor. The new boundary conditions for AdS$_2$ can be applied to this spacetime under which the Lie brackets of the asymptotic Killing vectors form a Virasoro and $U(1)$ Kac-Moody algebra without central extension. The central terms can be recovered by considering the Dirac brackets of the linearized covariant charges associated to the asymptotic Killing vectors at the boundary within the near horizon region. In the non-extremal case, the specific vector fields are determined by the diffeomorphisms that keep the Casimir operator as well as the $U(1)$ generator for the global warped conformal symmetry invariant. The selected vector fields thus will also keep the scalar radial equation and its frequency in the near horizon region invariant. The Lie brackets of these vector fields also satisfy the classical warped conformal symmetry. The central terms in the non-extremal case can be recovered by considering the Dirac brackets of the linearized covariant charges associated to these vector fields at the horizon bifurcation surface with a proper counterterm \eqref{ct}. The counterterm here is introduced for eliminating the mixed central extension term in deriving the standard warped conformal algebra. Our counterterm works for a large class of  spacetime geometries that endowed with metric expansions near their horizon bifurcation surfaces in the form of \eqref{nbifg}, since the cancellation of the mixed central extension term does not depend on the details of the functions \eqref{FF}.

The central terms in the standard warped conformal algebra obtained from the extremal and non-extremal cases are consistent with each other. In the extremal limit, the Virasoro central charges calculated by the two methods are the same and the Kac-Moody levels from the two cases match when the finite periodicity for the Bondi-like time coordinate $u$ mentioned in \eqref{tau} is $2\pi i$. Using the algebra as well as the central extension terms of the warped conformal symmetries derived in the bulk spacetime, the entropy formula of a finite temperature warped CFT which only depends on the symmetries, central terms, and modular properties of the theory then can be used to reproduce the horizon entropy with a precise agreement.

\section*{Acknowledgement}
We are grateful to Hai Lin, Jianxin Lu, and Wei Song for helpful discussions. JX is supported by the NSFC Grant No. 12105045.

\end{document}